# Adding artificial noise for code rate matching in continuous-variable quantum key distribution

*Sören Kreinberg, Igor Koltchanov, André Richter*
*VPIphotonics GmbH, Carnotstr. 6, 10587 Berlin, Germany*
*e-mail: soeren.kreinberg@vpiphotonics.com*

The reconciliation step of continuous-variable quantum key distribution protocols usually involves forward error correction codes. Matching the code rate and the signal-to-noise ratio (SNR) of the quantum channel is required to achieve the high reconciliation efficiencies that are crucial for long distance links. Puncturing and shortening is a way to adapt the code rate to the SNR at the cost of a slightly reduced reconciliation efficiencies. Instead of adapting the code rate to the SNR, we propose to add a controlled amount of artificial noise to the measured data, so that the resulting SNR could be reduced to match the given code rate. We show that our method can compete with puncturing and shortening and even outperform it in high-loss, high-excess noise scenarios.

## Introduction

In continuous-variable QKD [1], Alice and Bob are provided with correlated analogue data from the quantum channel. This correlation provides mutual information $I_{AB}$ to Alice and Bob. During an information reconciliation step, they exploit this mutual information to obtain a common bit sequence. A common approach is to use mathematical transformations to map their correlated noise to a virtual binary channel with additive white gaussian noise (BAWGNC). A simple option for the low-SNR regime is multidimensional reconciliation [2], [3], another important option is slice reconciliation, which also works for higher SNRs [4], [5]. Both reconciliation protocols use soft-decision forward error correction (FEC) codes to obtain an error-free bit sequence from the noisy virtual binary channel. In continuous-variable QKD, the mutual information shared by Alice and Bob $I_{AB}$ is often only slightly larger than the upper bound of the information an attacker Eve could have obtained (the so-called Holevo bound $\chi$). An efficient exploitation of the mutual information is therefore crucial for a secret key exchange: If the bit sequence held by Alice and Bob after reconciliation is shorter than the Holevo bound on Eves information, the extraction of a secret key is impossible. The length of the secret key depends on the secret fraction per transmitted symbol $r$ which is defined via the number of bits reconciled per symbol $b$ and the Holevo bound $\chi$ on Eves information (also per symbol) as $r = b - \chi$. Consequently, $b = \beta I_{AB}$ should be as high as possible for a given mutual information $I_{AB}$ (per symbol), and hence the reconciliation efficiency $\beta$ should be maximized. In practice, ideal efficiency $\beta = 1$ cannot be achieved, because information is lost during the mapping from the quantum channel to the BAWGNC, or because the applied FEC requires a higher SNR (i.e. more mutual information) to correct a given number of payload bits than theoretically necessary. Most forward error correction codes of a given length $N$ can transport only a fixed number of bits $k$. They are said to have the code rate $R = \frac{k}{N}$. To achieve a given frame error rate, they require a certain minimum mutual information $I_{\text{BAWGNC,min}} = R/\beta_{\text{FEC,max}}$. If the mutual information $I_{\text{BAWGNC}}$ is lower than $I_{\text{BAWGNC,min}}$, the desired frame error rate cannot be achieved. If the mutual information is higher than this value, the number of transported bits is still the same $b = \beta I_{AB} \propto R = const$. Consequently, the reconciliation efficiency $\beta = b/I_{AB} \leq R/I_{\text{BAWGNC}}$ drops. The Holevo bound $\chi$ on the other hand increases with higher SNR, reducing the secret fraction $r = b - \chi$. Therefore, it is important to have the ability to match code rate and mutual information (which is strongly linked to the SNR). Because of possibly unpredictable variations in the quality of the quantum channel, a-priori matching the signal power or mixing analogue noise into the signal is not sufficient. Instead, the ability for code rate matching during post processing is required. Via puncturing and shortening [6], [7] it is possible to change the code rate. The applicability of this approach to CV-QKD has been demonstrated by Wang et al. [8]. However, puncturing and shortening comes at the cost of a reduction of the reconciliation efficiency $\beta$ by more than half a percent. While this is acceptable on short distances, it can make a secure key exchange impossible on long distances. Here we present the opposite approach to match code rate and mutual information: Instead of changing the code rate, we artificially reduce the mutual information.

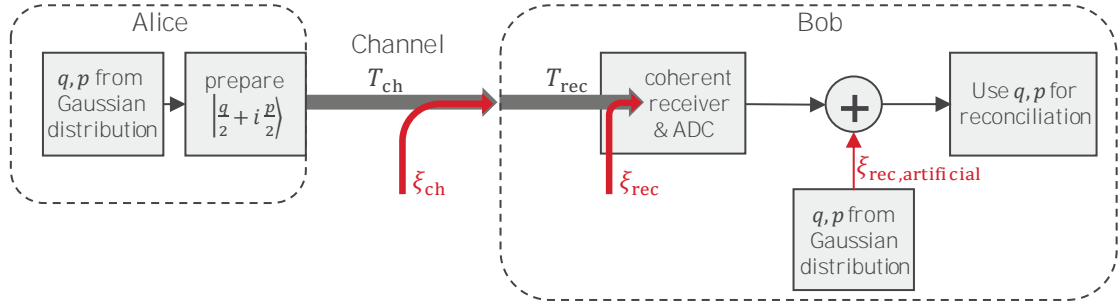

Figure 1 *Schematic of the QKD setup. Alice samples quadratures q,p from a Gaussian distribution with variance $V_{mod}$, prepares corresponding coherent states and sends them over the optical channel to Bob. The channel has a transmittance $T_{ch}$ and adds excess noise $\xi_{ch}$. Coupling losses on Bob's side and the limited quantum efficiency of his receiver is described by $T_{rec}$. During detection, Bob's receiver adds excess noise $\xi_{rec}$ to the signal. In the digital domain, Bob has the option to add extra noise $\xi_{rec,artificial}$ to the measured quadratures before using them for reconciliation. By doing so he matches the SNR to the FEC code used during reconciliation.*

## Results

We discuss our approach on the example of prepare-and-measure Gaussian-modulated CV-QKD [9] using reverse reconciliation [10]. We assume that both quadratures are measured simultaneously via two balanced homodyne detectors [11], often called "heterodyne detection" in the QKD community [12]. A schematic of the QKD setup is shown in Fig. 1. Very recently, it was calculated how trusting the receiver noise and receiver quantum efficiency reduces the Holevo bound and thus increases the secret fraction [13]. Most interesting, it was found that in scenarios with bad channel transmittance and high channel noise, it can be beneficial to add noise on Bobs side [13], [14]. We use this effect to match the mutual information to the code rate. For this purpose, Bob deliberately adds digital Gaussian white noise to his measurement results during post processing. Using the formalism derived by Laudenbach and Pacher [13], we calculate the secret fraction for different channel excess noise $\xi_{ch}$. We assume that multidimensional reconciliation and a multi-edge-type LDPC code of rate $R = 0.1$ is used, yielding a reconciliation of efficiency $\beta = 93.95\ \%$ [6], [8]. This implies that the reconciliation via this LDPC code requires a minimum SNR of $SNR_{R=0.1} = 2^{2R/\beta} - 1 = 0.159$. We further assume that we can trust the calibration of Bob's receiver with a trusted efficiency of $T_{rec} = 0.7$ and a trusted noise of $\xi_{rec} = 0.01$. For each value of the channel excess noise $\xi_{ch}$, the modulation variance $V_{mod}$ was optimized to achieve maximum secret fraction $r$ under the constraint that the resulting SNR matches the capacity of the LDPC code. Since excess noise, modulation variance, transmittance and SNR are linked via $SNR = \frac{T\ V_{mod}}{2+\xi_{ch}+\xi_{rec}}$, optimizing $V_{mod}$ for fixed SNR and excess noise also yields a corresponding optimum $T = T_{ch}T_{rec}$. These optimum parameter sets are depicted as circles in Fig. 2. Keeping these parameter sets, only the channel transmittance was changed. Under the assumption that a code with the same reconciliation efficiency $\beta = 93.95\ \%$ is available for arbitrary code rates (i.e. for arbitrary SNRs), the secret fractions corresponding to the pale dashed lines can be achieved. On the other hand, if the same $R = 0.1$ code is always used for all SNRs without any attempts of matching SNR and code rate, much smaller secret fractions (pale lines) can be achieved. This drastic secret fraction penalty inflicted by non-matched code rates can be partly mitigated by shortening and puncturing [8]. However, the reconciliation efficiencies for shortened and punctured codes are lower than for the original code. The secret fractions corresponding to the values presented in ref. [8] are depicted as crosses. Instead of applying puncturing and shortening, Bob can add additional Gaussian white noise $\xi_{\text{rec,artificial}} = \frac{TV_{\text{mod}}}{SNR_{R=0.1}} - 2 - \xi_{\text{ch}} - \xi_{\text{rec}}$ to his data. Thus, he can reduce the SNR and the mutual information between him and Alice to match the available reconciliation scheme and LDPC code. This approach is beneficial compared to directly using the available reconciliation scheme without any mitigation, because adding noise not only reduces the SNR, but also reduces the Holevo bound on Eve's information $\chi$. It was pointed out recently, that deliberately adding trusted noise can even increase the secret fraction in some scenarios [13]. This is also reflected by the calculations (solid lines). In most scenarios, adding artificial noise to Bob's measurements can compete with the puncturing and shortening approach. Under delicate low-secret-fraction conditions, our approach delivers much higher secret fractions than the puncturing and shortening approach, because puncturing and shortening slightly reduces the reconciliation efficiency $\beta$ by approximately 0.5 % [8].

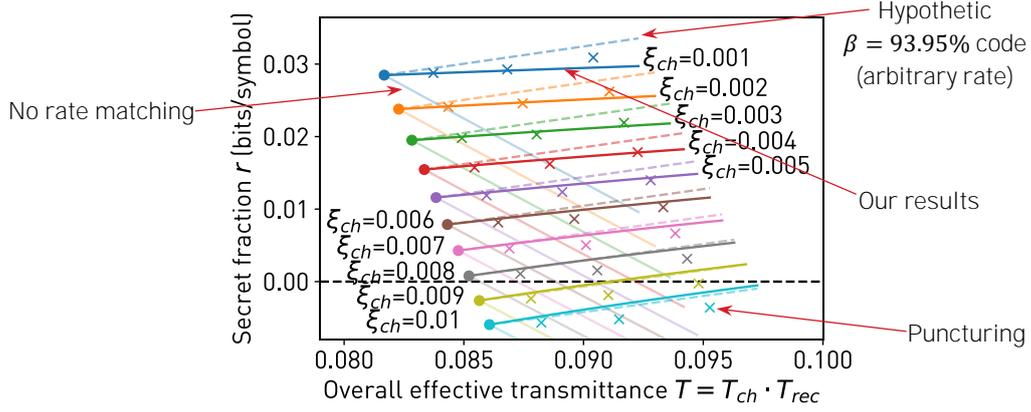

Figure 2 *Mitigation of transmittance increase-induced secret fraction drop for different channel excess noise values $\xi_{ch}$. The circles correspond to the optimum parameter set which yield the nominal SNR (0.159) of the $\beta = 93.95\%, R = 0.1$ code. The curves depict the secret fraction that can be achieved without rate-matching (pale solid lines), with a code of arbitrary rate and $\beta = 93.95\%$ (pale dashed line), by adding noise for matching the mutual information to the code rate (solid lines). The crosses correspond to the results obtained by Wang et al. [8] using puncturing.*

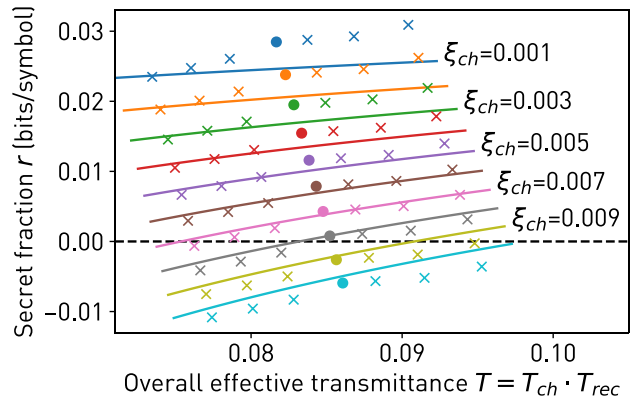

Figure 3 *Code rate matching. The circles correspond to the optimum parameter set which yield the nominal SNR (0.159) of the $\beta = 93.95\%, R = 0.1$ code. The crosses correspond to the results obtained by Wang et al. [8] using puncturing and shorting. The curves depict the secret fraction that can be achieved when operating with an 15 %-increased $V_{\text{mod}}$ and when adding noise in order to match the mutual information to the code rate.*

Up to here, our approach only mitigates transmittance *increases* whereas a *decrease* in transmittance would increase the frame error rate, ultimately rendering reconciliation impossible. Therefore, artificial noise needs to be added already at the operation point. Thus, if the transmittance decreases, rate-matching can be achieved by reducing the artificial noise. In order to be able to add artificial noise at the channel parameters corresponding to the operation point of the given scenario, we increase the modulation variance by 15 %. The resulting secret fraction is shown in Fig. 3 alongside the secret fraction that would result from the puncturing and shortening reconciliation efficiencies published by Wang et al. [8]. For low channel excess noise, puncturing and shortening yields higher secret fractions than our proposed method. However, when the channel excess noise is high and the secret fraction is low, rate-matching by adjusting the amount of artificial noise outperforms puncturing and shortening.

## Conclusion

Using prepare-and-measure weak coherent Gaussian-modulated CV-QKD as an example, we demonstrated that matching the mutual information to the FEC code rate by adding artificial noise can compete with code rate-adaption via puncturing and shortening. Especially when the channel is noisy, the proposed method outperforms puncturing and shortening. This advantage comes partly from the reconciliation efficiency penalty inflicted by puncturing and shortening [8] and partly from the effect, that adding noise can actually increase the secret key rate under certain conditions [13].

## Acknowledgement

This work has received funding from the European Union's Horizon 2020 research and innovation programme through the Quantum-Flagship project UNIQORN under grant agreement No 820474.